# Thermoelectric microscopy of magnetic skyrmions

Ryo Iguchi[1], Shinya Kasai[1,2], Kazushige Koshikawa[3], Norimichi Chinone[3], Shinsuke Suzuki[3], and Ken-ichi Uchida[1,4,5]

[1]National Institute for Materials Science, Tsukuba 305-0047, Japan.

[2]PRESTO, Japan Science and Technology Agency, Saitama 332-0012, Japan.

[3]Hamamatsu Photonics K.K., Hamamatsu 431-3196, Japan.

[4]Department of Mechanical Engineering, The University of Tokyo, Tokyo 113-8656, Japan.

[5]Center for Spintronics Research Network, Tohoku University, Sendai 980-8577, Japan.

Correspondence and requests for materials should be addressed to R.I. (e-mail: IGUCHI.Ryo@nims.go.jp)

**The magnetic skyrmion is a nanoscale topological object characterized by the winding of magnetic moments, appearing in magnetic materials with broken inversion symmetry. Because of its low current threshold for driving the skyrmion motion, they have been intensely studied toward novel storage applications by using electron-beam, X-ray, and visible light microscopies. Here, we demonstrate another imaging method for skyrmions by using spin-caloritronic phenomena, that is, the spin Seebeck and anomalous Nernst effects, as a probe of magnetic texture. We scanned a focused heating spot on a Hall-cross shaped MgO/CoFeB/Ta/W multilayer film and mapped the magnitude as well as the direction of the resultant thermoelectric current due to the spin-caloritronic phenomena. Our experimental and calculation reveal that the characteristic patterns in the thermoelectric signal distribution reflect the skyrmions' magnetic texture. The thermoelectric microscopy will be a complementary and useful imaging technique for the development of skyrmion devices owing to the unique symmetry of the spin-caloritronic phenomena.**

The antisymmetric exchange interaction, Dzyaloshinskii-Moriya interaction, mediated by spin—orbit interaction induces a swirling magnetic object called the magnetic skyrmion in non-centrosymmetric ferromagnets[1–5] and magnetic multilayer systems[6–10]. Since the real-space observation of the magnetic skyrmions[3], they have attracted substantial attention from the viewpoints of not only fundamental physics but also spintronic device applications[11–13]. The skyrmions are expected to realize energy-efficient non-volatile memories because of the high stability of their magnetic texture and the low charge current density for driving their motion, which can be several orders of magnitude smaller than the current density for driving conventional magnetic objects[6,7,14]. The characteristic of the skyrmions is their internal magnetic texture; the magnetic moments in the skyrmions are twisted by Dzyaloshinskii-Moriya interaction and their directions cover a whole surface of a sphere, leading to the non-trivial winding number[15,16].

To date, the emergence and dynamics of the magnetic skyrmions have been intensively studied in real space by using a variety of microscopy methods. For instance, Lorentz-transmission and spin-polarized low-energy electron microscopies reveal the internal texture of Bloch-[3,17], Néel-[18], and anti-skyrmions[4]. The spin-polarized scanning tunnelling microscopy reveals the atomic-scale skyrmions[19]. The magneto-optical, magnetic-force, and X-ray microscopies are used for detecting the position and dynamics of the skyrmions[6,8,10,20–23], eventually reaching the observation of the skyrmion Hall effect[24,25].

In this work, we demonstrate another microscopy method for observing the skyrmions: thermoelectric microscopy of the skyrmions realized by spin-caloritronic phenomena (Fig. 1a). The spin-caloritronic phenomena refer to the thermoelectric or thermo-spin effects in which the conversion property between heat and charge currents depends on the magnetization direction in magnetic systems[26–29]. When a local temperature gradient is generated by focused heating and its spot size is smaller than the diameter of skyrmions, the spin-caloritronic signals reflect the local variation of magnetic moment direction in the skyrmions. A representative example of such spin-caloritronic phenomena is the anomalous Nernst effect (ANE) in ferromagnetic conductors, which is the thermoelectric analogue of the anomalous Hall effect[27,28]. The ANE induces a charge current $\mathbf{j}_c$ along the outer product of the applied temperature gradient $\nabla T$ and the magnetic moment $\mathbf{m}$:

$$\mathbf{j}_c \propto \mathbf{m} \times \nabla T \tag{1}$$

When the local heating spot is scanned, the ANE-induced $\mathbf{j}_c$ changes its direction with the local $\mathbf{m}$ and $\nabla T$ distributions; for film samples, the out-of-plane (in-plane) $\nabla T$ induces $\mathbf{j}_c$ in response to the in-plane (out-of-plane) $\mathbf{m}$ (Fig. 1b). The resultant thermoelectric image of the $\mathbf{j}_c$ direction exhibits the quarter-turned in-plane texture (curl of the out-of-plane moment vector) of magnetic skyrmions for the out-of-plane (in-plane) $\nabla T$ (Fig. 1c). Another example is the magnon-driven or electron-driven spin Seebeck effect (SSE)[30,31]. The SSE has a chance for exclusively probing the in-plane magnetic texture of skyrmions in combination with the inverse spin Hall effect because the SSE-induced $\mathbf{j}_c$ is generated only by the out-of-plane $\nabla T$ across ferromagnet/paramagnet junctions owing to the following symmetry[27–29,32,33]:

$$\mathbf{j}_c \propto \begin{cases} \mathbf{m} \times \nabla T & \text{for } \nabla T || \mathbf{n} \\ 0 & \text{for } \nabla T \perp \mathbf{n} \end{cases} \tag{2}$$

where $\mathbf{n}$ denotes the interface normal of the junction (Fig. 1b). The thermoelectric microscopy based on the spin-caloritronic phenomena will be useful for studying skyrmion physics and its device applications because of the usability of heat; thanks to heat propagation, this method is applicable to skyrmions in magnetic layers embedded in protection or passivation layers when those layers are electrically insulated from the magnetic layers and thin enough to keep high spatial resolution. This capability will be useful for evaluating skyrmion devices without complicated pre-processes, such as removing protection or passivation layers.

**Results**

**Thermoelectric imaging of skyrmions.** In the experiment, we used a Hall cross-shaped MgO/CoFeB/Ta/W film, where Néel skyrmions appear owing to the interfacial Dzyaloshinskii-Moriya interaction and perpendicular magnetic anisotropy[34]. For the local heating, we applied a focused laser beam to the sample by using an ultraviolet laser with a wavelength of 405 nm to achieve finer optical resolution than that due to visible light[35–37]. The beam waist of the focused laser spot is expected to be about 440 nm in the diameter (see Supplementary Fig. S1 for optical performance). To obtain thermoelectric images, the focused laser spot was scanned in a 16.8×16.8 μm$^2$ area of the *x-y* plane with 512×512 points and, at each laser position, the generated thermoelectric current was measured with a current meter (Fig. 2a).

Figure 2b shows the optical reflection and thermoelectric images, where the thermoelectric current in the *x*-direction $j_c^x$ was measured under a perpendicular field $\mathbf{H}_z$ (with the magnitude $H_z$). The thermoelectric image, as well as the optical reflection image, shows no contrast inside the Hall cross at $\mu_0 H_z$ = -10.9 mT with $\mu_0$ being the vacuum permeability, where the magnetization of the CoFeB layer is fully aligned along the $\mathbf{H}_z$ direction. When $\mu_0 H_z$ is increased to -1.6 mT, the thermoelectric image exhibits patchy patterns inside the Hall cross while the optical image remains unchanged; this is the signature of the spin-caloritronic phenomena reflecting the magnetic texture (for magneto-optical images, see Supplementary Fig. S2). Note that the dark and white feature at the edges of the Hall cross in the thermoelectric images may be attributed to the ANE contribution caused by the asymmetric temperature gradient at edges (see Supplementary Information for detail)[38]. By further increasing $H_z$, the thermoelectric images exhibit various patterns. We found that small isolated circular objects are generated at $\mu_0 H_z$ = +4.0 mT, where the magnetic domains are expected to be decomposed into skyrmions[34] (see Supplementary Fig. S2 for correspondence with the magnetic texture imaged by the magneto-optical Kerr effect). The circular objects are wiped out from the thermoelectric image when $H_z$ becomes larger than the saturation field for the CoFeB layer. Importantly, all the circular objects in the thermoelectric image comprise both positive and negative intensities respectively at the upper and lower halves of the objects. This is consistent with the expected behaviour for the skyrmions (see the $x$ component of $\mathbf{j}_c$ distribution in Fig. 1c), demonstrating the thermoelectric detection of skyrmions.

**Mapping of thermoelectric current direction and magnitude with Hall cross structure.** The thermoelectric microscopy allows us to visualize not only the magnitude but also the direction of the thermoelectric current simply by measuring the signals in the two orthogonal directions of the Hall cross. The obtainable spatial information is linked to the skyrmions' in-plane and out-of-plane texture depending on the thermal condition (as discussed later). We measured the thermoelectric images of $j_c^x$ and $j_c^y$, and calculated the thermoelectric current vector $\mathbf{j}_c = j_c^x \mathbf{x} + j_c^y \mathbf{y}$ with $\mathbf{x}$ ($\mathbf{y}$) being the unit vector along the $x$ ($y$) direction. Figure 3a shows the total amplitude image and the pseudo-colour image representing the $\mathbf{j}_c$ direction (note that only the centre area of the Hall cross is extracted to ensure the same sensitivity for $j_c^x$ and $j_c^y$). Figure 3b shows the evolution of the synthesized thermoelectric images with respect to $H_z$, where the field values are shown in the magnetization curve in Fig. 3c with the alphabetical labels A-L. The images at B -F, and at G-K of Fig. 3b show the magnetic domains and skyrmions, respectively. As shown in Fig. 3d, the thermoelectric patterns due to the skyrmions show that the **j**$_c$ direction continuously rotates and covers 360° in the *x-y* plane with the same chirality. The two-dimensional thermoelectric detection can enrich the information about the spatial distribution.

**Simulation of thermoelectric images due to skyrmions.** Next, we analyse the thermoelectric images due to the skyrmions. The focused laser heating induces not only the out-of-plane but also in-plane temperature gradients; these contributions respectively reflects the in-plane (with its vector $\mathbf{m}_{||} = m_x \mathbf{x} + m_y \mathbf{y}$) and out-of-plane (with its component $m_z$) magnetic moments, and thus have to be distinguished from each other to understand the relation between the observed thermoelectric images and the magnetic textures. To do this, the temperature profile in the model shown in Fig. 4a is calculated in a cylindrical coordinate (*r* and *z* for each axis) using a finite element method[39], where the laser heating with a power of 0.4 mW (considering the reflection) and a heating spot diameter of 440 nm is assumed. Figure 4b shows the expected temperature profile and the resultant out-of-plane and in-plane temperature gradients ($\partial_z T$ and $\partial_r T$, respectively). By convoluting the $\partial_z T$ and $\partial_r T$ distributions and the magnetic texture of the Néel skyrmions (shown in Fig. 4c), the **j**$_c$ distributions induced by the ANE and SSE are calculated using equations (1) and (2), respectively. As shown in Fig. 4d, the ANE induces the clock-wise pattern of the **j**$_c$ direction either due to $\partial_z T$ or $\partial_r T$, while the SSE induces the counter-clock-wise pattern only due to $\partial_z T$. This difference between the ANE and SSE signals originates from the positive ANE coefficient in CoFeB and the negative spin–charge conversion in Ta and W for the SSE[32,40]. Here, the sign and magnitude of the ANE and SSE signals used for the calculations were determined experimentally (see Supplementary Information for details). The dominant contribution in our sample is found to be the $\partial_r T$-induced ANE, reflecting $m_z$, which can be confirmed by the chirality of **j**$_c$ consistent with the experimental results. We note that the conversion of our thermoelectric images to the magnetization distribution is impossible since the dimensions are different between **j**$_c$ (two: $j_c^x$ and $j_c^y$) and **m** (three: **m**$_{||}$ and $m_z$). Nevertheless, owing to the sensitivity to **m**$_{||}$, the thermoelectric images can differ between the Néel skyrmions and trivial magnetic bubbles (Fig. 4e)[41]; the calculation results show that the skyrmion induces the **j**$_c$ signal distributed uniformly over its circumference while the bubble induces that distributed non-uniformly (Fig. 4f). We note, however, that the experimental distinction in our experimental results is difficult because of low S/N ratio (see Supplementary Fig. S3 for circumferential profiles on the skyrmions). Another interesting point is that the **j**$_c$ direction steeply changes around the skyrmions' core as shown in Fig. 3d. The additional calculation on the skyrmion diameter dependence confirms that the steep change remains within the heating spot diameter even when the skyrmion diameter is comparable to the heating spot diameter (see Supplementary Fig. S4). This is the characteristic of the thermoelectric generation due to the spin-caloritronic phenomena; the polarity of **j**$_c$ changes when the heating spot crosses the skyrmions' centre. These facts confirm that the spin-caloritronic phenomena in magnetic systems can be a probe for the skyrmions.

The difference in the thermoelectric images of the skyrmions and trivial bubbles can be made more pronounced by suppressing the $\partial_r T$ ($\mathbf{m}_\parallel$) contribution and increasing the $\partial_r T$ ($m_z$) contribution. This can be achieved by changing the thermal design of the sample system, in particular, the thermal resistance of the layers beneath the magnetic layer hosting the skyrmions. Here, we show how the thermoelectric images change with the thermal condition by calculating the dependence of the ANE signals on the substrate thermal conductivity $\kappa_{\text{subs}}$ for the sample layer directly located on the substrate. The right panel of Fig. 5a shows that, as $\kappa_{\text{subs}}$ increases, the magnitude of $\partial_z T$ ($\partial_r T$) increases (decreases). This behaviour can be understood in terms of the thermal resistance from the surface of the sample to the bottom. When $\kappa_{\text{subs}}$ is low, to reduce the thermal resistance, the cross-sectional area of the heat flux is expanded, which results in the increased $\partial_r T$ component compared with the high $\kappa_{\text{subs}}$ case. Accordingly, the resultant thermoelectric images at the improved $\kappa_{\text{subs}}$ value are governed by the $\partial_z T$ contribution as shown in Fig. 5b. In this condition, the clear difference between the skyrmions and trivial bubbles can be obtained because they differ in the $\mathbf{m}_\parallel$ distribution and the $\partial_z T$ contribution reflects $\mathbf{m}_\parallel$ (see the $\partial_z T$-induced images in Fig. 4d,f). The improvement monotonically continues with increasing $\kappa_{\text{subs}}$, in the range of $\kappa_{\text{subs}}$ investigated here, as shown in Fig. 5c, demonstrating the importance of the thermal design of the system for enhancing the $\mathbf{m}_\parallel$ sensitivity. When the spatial resolution and the $\mathbf{m}_\parallel$ sensitivity are sufficiently large for disregarding the $m_\perp$ contribution, the thermoelectric images can be converted into the $\mathbf{m}_\parallel$ distribution, which can be used to distinguish magnetic objects characterized by the $\mathbf{m}_\parallel$ configuration, i.e. not only Néel skyrmions and trivial bubbles but also Bloch skyrmions. We also note that in the model for Fig. 4, the oxidized Si layer acts as the substrate and its low thermal conductivity diminishes the $\partial_z T$ contribution.

**Application of thermoelectric microscopy to current-induced skyrmion transport.** Importantly, the thermoelectric microscopy is applicable directly to investigations of skyrmion transport since the electrodes used for the thermoelectric detection is available also for the charge current application to drive the skyrmion motion. Figure 6a shows a schematic of the thermoelectric imaging combined with charge-current pulse application to the same sample. The pulse amplitude is about 9 mA, the charge current density of which is $0.4 \times 10^{12}$ A/m$^2$, and the pulse duration is 100 ns. The thermoelectric images are recorded at the fields with the labels A-C in Fig. 6b, where at A skyrmions appear and at B and C domains. Note that the field polarization is reversed from Fig. 2b and the signal contrast is also reversed. The images at A of Fig. 6c shows that the skyrmions maintain its form after applying the charge current pulses as expected from the energy barrier related to the topological texture. The motion of the individual skyrmions is difficult to be identified from the evolution of the thermoelectric images because of the presence of many skyrmions and/or the thermal activation due to pulse application. The images at B and C of Fig. 6c show the nucleation rather than the motion, showing the different dynamic properties between domains and skyrmions. The experiment demonstrates the feasibility for measuring skyrmion transport, confirming the versatility of the proposed imaging method.

**Discussion**

The advantages of the thermoelectric microscopy are the usability owing to heat and sensitivity to in-plane magnetic moments. Since the heat can propagate in materials differently from conventional optical or electron-beam probes, the thermoelectric microscopy can expand the scope of target material systems. For example, it may be applicable to the systems where the magnetic layers are covered by additional layers[42,43] and the conventional optical or electron-beam probes cannot access the skyrmions. Note that the electrical insulation and low thermal resistance ensure sufficient spatial resolution. Thus, the proposed method will be favourable for finding new forms and functions of skyrmion devices with cap layers and for evaluating the performance of skyrmion devices without complicated pre-processes for removing over layers and/or without structural limitations for transmitting probe beams. This advantage makes the thermoelectric microscopy fascinating along with the capability of the measurements at atmospheric pressure with table-top equipment. The in-plane magnetic moment sensitivity of the thermoelectric microscopy would be useful for the fundamental science to probe the internal texture of the skyrmion families[4,5,44,45]. Although the experimental results reported here reflect the out-of-plane magnetic moments, the situation can be changed by optimizing the thermal and/or thermoelectric configuration of the system to enhance the $\partial_z T$ and/or SSE contributions as discussed above. The thermoelectric microscopy does not require the change of the measurement configuration for full angular resolution of the in-plane **m** distribution; it is achieved simply by measuring the thermoelectric currents in the two orthogonal directions owing to the transverse symmetry of the ANE and SSE. This is substantially different from the Lorentz transmission electron-beam and X-ray microscopies, which require the oblique incidence and the rotation of the beam or samples[8,46].

The thermoelectric microscopy based on the spin-caloritronic phenomena creates the following synergies in skyrmion studies. Most of the skyrmion devices are based on a thin film form with electrodes for applying a driving charge current, and thus ready for the thermoelectric microscopy. In addition, materials used for the skyrmion devices can exhibit large thermoelectric or thermo-spin conversion because the conversion efficiency is determined by the strength of the spin—orbit interaction and it is also the source for Dzyaloshinskii-Moriya interaction necessary to stabilize the magnetic moments in the skyrmions. A possible drawback of the laser heating is the spatial resolution (followed by the optical diffraction limit), but this could be resolved by using near-field optics[38,47], such as scanning near field optical microscopy, or another energy flux, such as electron beams[48], for focused heating. If the spatial resolution of the thermoelectric microscopy is further improved, the real-space exploration of the topological thermoelectric signals due to the emergent field of the skyrmions[49] comes in scope, which will be an interesting topic in spintronics and thermoelectrics. It is also intriguing to study the temporal response of the thermoelectric current because the difference in the activation time of the thermoelectric signals may give a clue to find collective dynamics in skyrmions. We thus anticipate that the thermoelectric microscopy becomes a useful technique for developments in physics and applications of skyrmions.

**Methods**

**Sample fabrication.** The film stack used in the measurements is Ta(3)/MgO(1.8)/Co$_{20}$Fe$_{60}$B$_{20}$(1.3)/Ta(0.4)/W(3) (unit in nm) deposited on a thermally oxidized Si substrate. Post annealing in vacuum was applied at 300 °C for 30 min to enhance the interface induced perpendicular magnetic anisotropy. For obtaining angular information of the induced thermoelectric signals, the sample was patterned into a Hall cross structure with a bar width of 5 μm by using electron-beam lithography (shown in Supplementary Fig. S5). We note that our sample shares the same stack as the sample used in the skyrmion study[34].

**Thermoelectric imaging.** We employed an ultraviolet laser of which wavelength is 405 nm. The laser light was focused on the sample surface through an objective lens with the numerical aperture (NA) of 0.5. The expected beam waist ω is about 220 nm, where $\omega = \lambda(\pi\tan\theta)^{-1}$ and $NA = \sin\theta$ are used. The focused laser beam spot was scanned in a 16.8×16.8 μm$^2$ area with 32.8-nm point-to-point separation. The measurement duration for each point was 128 μs. After obtaining raw thermoelectric images, periodic environmental noise and uniform signal offset are eliminated from the images (see Supplementary Information and Supplementary Fig. S6). The $j_c^x$ and $j_c^y$ images are taken one after another by changing cable connections because the equipment used in this study has only one current meter. The different cable configurations may affect the noise level, which was different between $j_c^x$ and $j_c^y$ images. The laser power was set to 0.84 mW, which is low enough to keep the magnetic texture (see Supplementary Fig. S7). The incident light is circularly polarized and the magneto-circular dichroism (MCD) could be observed in our setup, but in the experiment, no contrast change appears in the optical images (Fig. 2b) because of the negligibly small MCD coefficient of the sample or the existence of the cap layer. The electrical connection to the sample was made by attaching tungsten needle probes to the Au electrodes located at the ends of the Hall cross. The perpendicular magnetic field was applied using an electromagnet. The measurements were performed at room temperature and atmospheric pressure.

**Simulation of thermoelectric images due to skyrmions.** The temperature profile in the model shown in Fig. 4a was calculated on the basis of the heat equation $\nabla \cdot [\kappa(z)\nabla T(r,z)] = 0$ in the cylindrical coordinate ($r$ and $z$ for radial and longitudinal axes, respectively) using a finite element method package[39].

The lateral length of the model (along the $r$ axis) is set to 2.5 μm and no azimuthal-angle $\phi$ dependence is assumed. The laser heating effect is treated as heat current injection from the top of the sample system, of which density is defined as,

$$j_{q,in} = P_{abs} \frac{2}{\pi \omega^2} \exp\left(-2\frac{r^2}{\omega^2}\right) \tag{3}$$

where ω is the beam waist in the radius (220 nm in our study). The absorbed power $P_{abs}$ is assumed to be 0.4 mW by considering the reflection (~50% of the applied laser power $P_{in}$ = 0.84 mW)[50], which may be lower in the experiment as the low thickness of the sample system allows transmission of the irradiated light. For simplicity in the simulation, the MgO/CoFeB/Ta/W multilayer and the cap layer are treated as a single layer (the sample layer) but the interfacial thermal resistance contribution[51] is incorporated as thermal conductivity anisotropy. After calculating the averaged temperature gradient profiles [$\overline{\partial_r T(r)}$ and $\overline{\partial_z T(r)}$] over the thickness direction in the sample layer (Fig. 4b), the thermoelectric responses were calculated using equations (1) and (2); the SSE contribution is calculated by $j_c^{x(y)} \propto +(-)\int m_{y(x)}(r,\phi)\overline{\partial_z T(r)}\,r\mathrm{d}r\mathrm{d}\phi$ and the ANE contribution is calculated by $j_c^{x(y)} \propto +(-)\int m_{y(x)}(r,\phi)\overline{\partial_z T(r)}r\mathrm{d}r\mathrm{d}\phi - (+)\int m_z(r,\phi)\overline{\partial_r T(r)}\sin\phi(\cos\phi)r\mathrm{d}r\mathrm{d}\phi$. Inside the skyrmions, $m_{x(y)}$ is assumed to change following a sinusoidal function with the frequency being the skyrmions' diameter of 1,000 nm (Fig. 4c). The chirality is assumed following the previous report on the CoFeB/Ta junction[6]. We note that the above calculation is for an idealised skyrmion and the deformation of the magnetic configuration due to skyrmions' size or fabricated structures affects the signal magnitude; the magnitude scales with the ratio of the area having the in-plane magnetic component or the gradient of the out-of-plane magnetization to the locally heated area. The relative magnitude of the ANE to the SSE is assumed to be 1.5 based on the conventional thermoelectric measurements in the out-of-plane and in-plane temperature-gradient geometries (see Supplementary Information for details)[52].

**Data availability.** The data that support the findings of this study are available from the corresponding authors upon reasonable request.

**Acknowledgments**

The authors thank Y. Hattori for valuable discussion and S. Sugimoto for technical assistance. This work was supported by CREST "Creation of Innovative Core Technologies for Nano-enabled Thermal Management" (JPMJCR17I1) and PRESTO "Topological Materials Science for Creation of Innovative Functions" (JPMJPR18L3) from JST, Japan; Grant-in-Aid for Scientific Research (S) (JP18H05246) and Grant-in-Aid for Early-Career Scientists (JP18K14116) from JSPS KAKENHI, Japan.

**Author Contributions**

R.I. and K.U. planned the study. R.I., S.K, and K.U designed the experiments. R.I., K.K., N.C., S.S., and K.U. performed the measurements and S.K prepared the sample. R.I. analysed the data and prepared the manuscript. R.I., S.K., K.K., N.C., S.S., and K.U. discussed the results, developed the explanation of the experiments, and commented on the manuscript.

**Additional Information**

**Competing Interests:** The authors declare no competing interests.

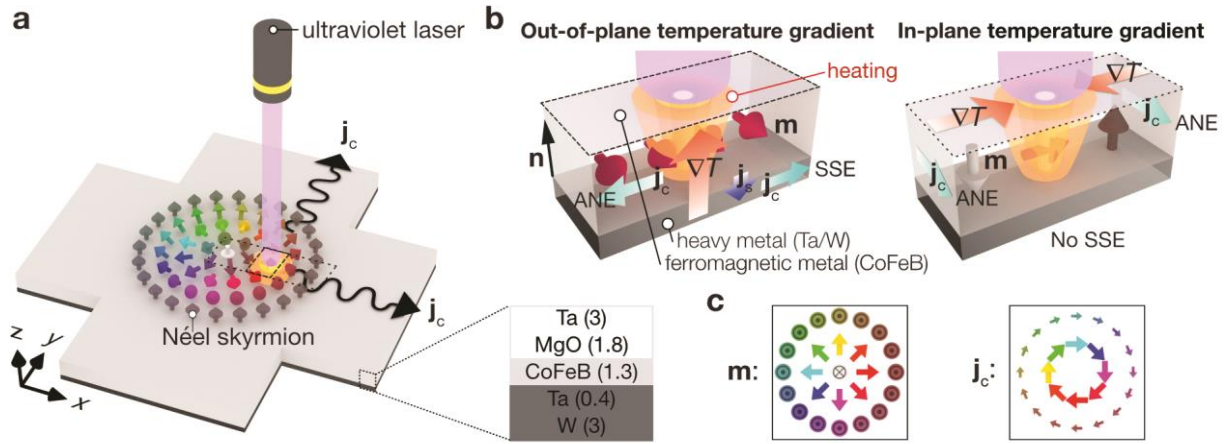

**Figure 1.** Thermoelectric microscopy of skyrmions. (**a**) Schematic illustration of the experimental setup. (**b**) Charge current generation due to spin-caloritronic phenomena by out-of-plane and in-plane temperature gradients $\nabla T$. The magnetic configurations which can induce non-zero signals coupled to the corresponding temperature gradients are shown. For the anomalous Nernst effect (ANE), $\nabla T$ induces the charge current (with its density vector $\mathbf{j}_c$) in the ferromagnetic layer. For the spin Seebeck effect (SSE), $\nabla T$ induces the spin current (with its density vector $\mathbf{j}_s$) along the interface normal $\mathbf{n}$ in the heavy metal layer and $\mathbf{j}_s$ is converted into the charge current via the inverse spin Hall effect. (**c**) Texture of magnetic moment $\mathbf{m}$ of a Néel skyrmion and expected $\mathbf{j}_c$ distribution due to the ANE and SSE induced by either out-of-plane or in-plane $\nabla T$.

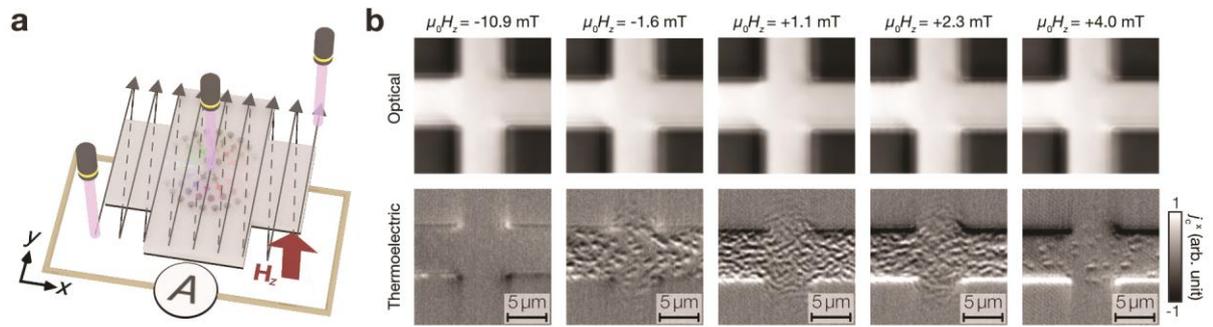

**Figure 2.** Thermoelectric detection of skyrmions. (**a**) Experimental configuration of the heating spot scanning and charge current measurement. (**b**) Evolution of optical (top) and thermoelectric (bottom) images with increasing the magnetic field value $H_z$. The optical images were taken by monitoring the reflection of the irradiated laser.

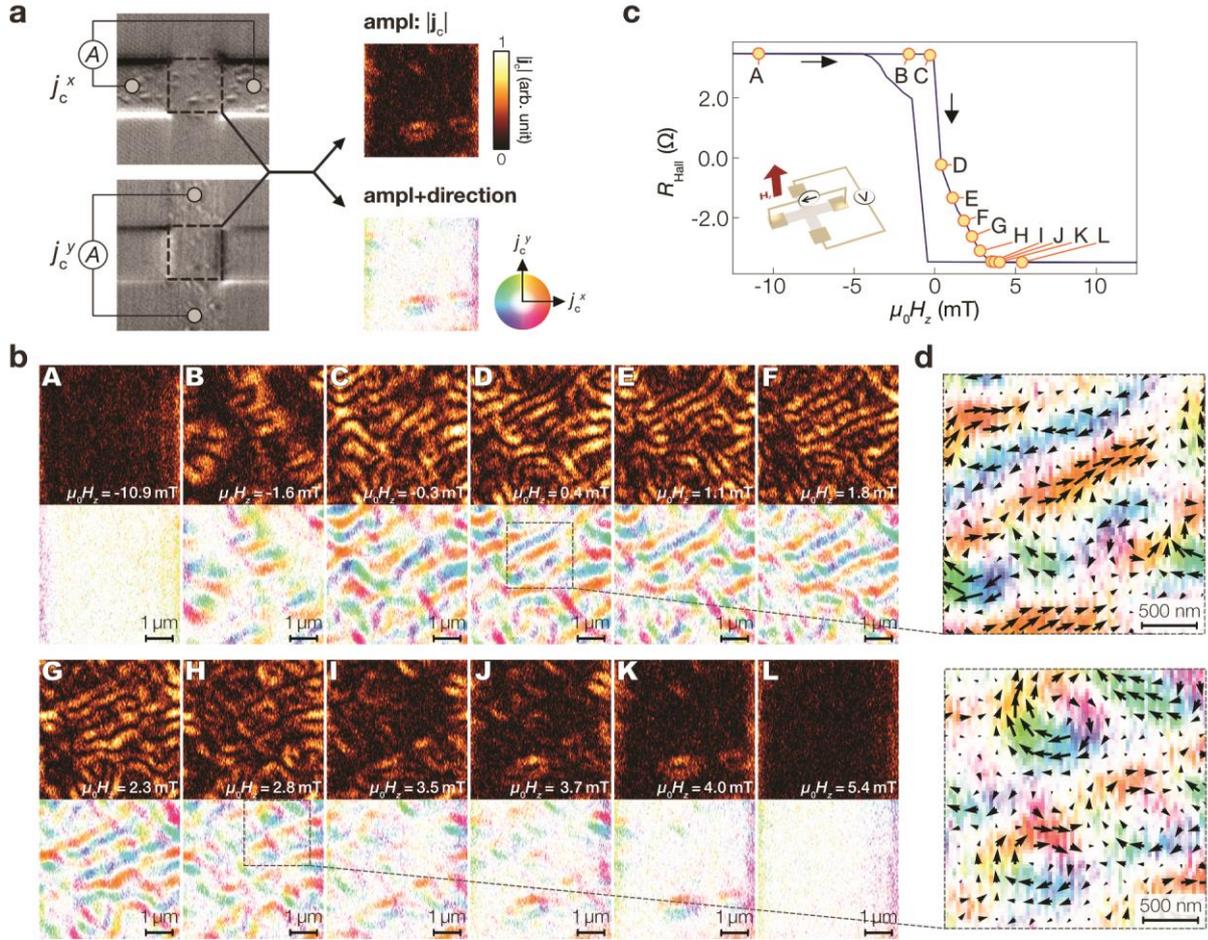

**Figure 3.** Two-dimensional thermoelectric detection of skyrmions. (a) Schematic of the synthesis of thermoelectric images. After $H_z$ was changed, $j_c^x$ and $j_c^y$ are recorded in order. Positional shift between the $j_c^x$ and $j_c^y$ images is compensated by comparing the optical reflection images recorded simultaneously with the thermoelectric images. (b) Synthesized thermoelectric images for various values of $H_z$. The top and bottom images respectively shows the ampltiude and the pseudo-color image reflecting amplitude and direction of $\mathbf{j}_c$. (c) $H_z$ dependence of the Hall resistance $R_{Hall}$ of the sample system, which is proportional to the magnetic moment perpendicular to the plane. The thermoelectric images were measured at the magnetic field marked with yellow circles in the $R_{Hall}$-$H_z$ curve. Note that $R_{Hall}$ measurements were performed

separately from the thermoelectric microscopy measurements. (**d**) Expanded view of the synthesized thermoelectric images with arrows indicating the direction of $\mathbf{j}_c$.

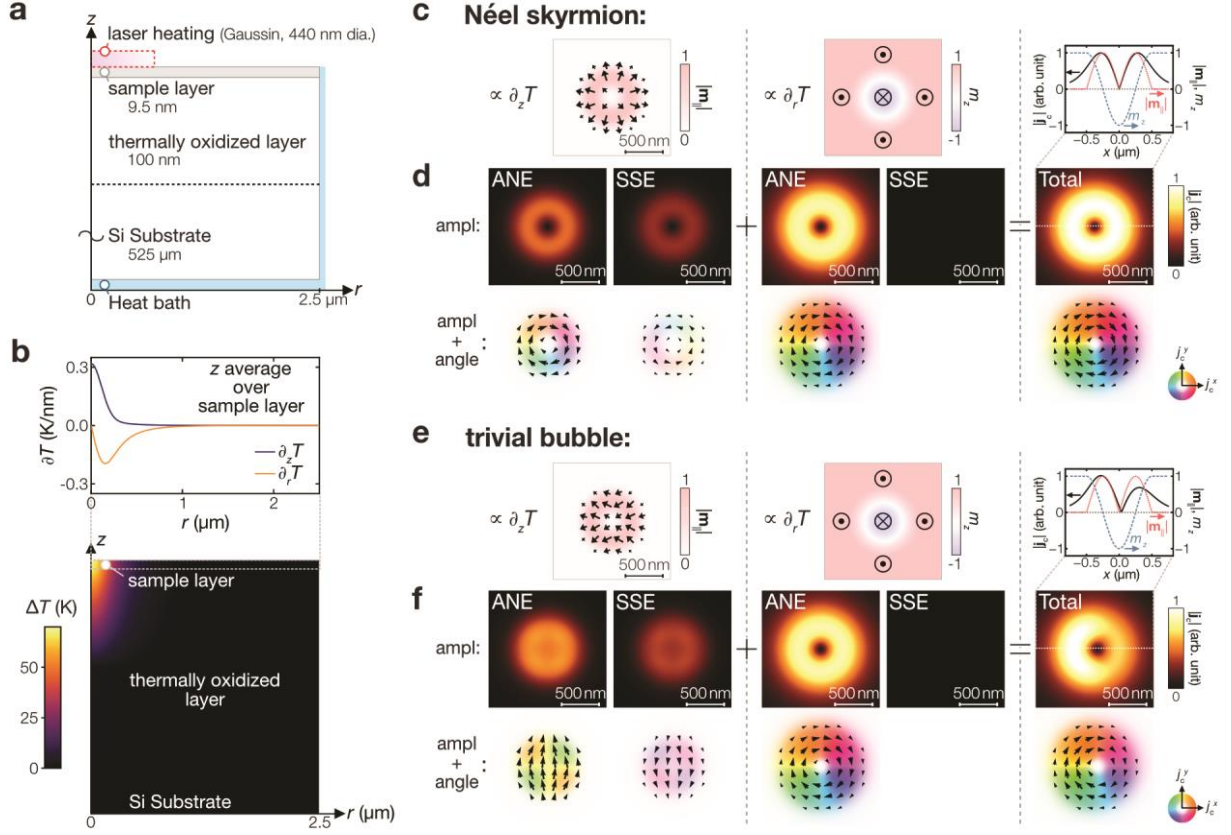

**Figure 4.** Calculation of thermoelectric images due to ANE and SSE. (**a**) Coordinate, dimensions, and parameters of the system used for temperature profile calculations. The thermal conductivity is assumed to be 10 (100) Wm$^{-1}$K$^{-1}$ for out-of-plane (in-plane) of the sample layer, 1.5 Wm$^{-1}$K$^{-1}$ for the thermally-oxidized layer, and 160 Wm$^{-1}$K$^{-1}$ for the Si substrate layer. (**b**) Calculated temperature profiles. The top panel shows the out-of-plane temperature gradient ($\partial_z T$) and in-plane temperature gradient ($\partial_r T$) in the sample layer, where the values are averaged along the $z$ axis in the layer. The bottom shows the temperature change $\Delta T$ around the sample layer. (**c,d**) Assumed magnetic components and resultant charge current signals for the Néel skyrmion with a diameter of 1 μm. (**e,f**) Assumed magnetic components and resultant charge current signals for the trivial bubble. The SSE and ANE induce the signals due to $\partial_z T$ and the out-of-plane magnetic moment with its magnitude $m_z$ while only the ANE induces the signals due to $\partial_r T$ and the in-plane magnetic moment $\mathbf{m}_\parallel$. The black arrows indicate the direction of the in-plane magnetic moments (c,e) and the induced $\mathbf{j}_c$ signal (d,f). The $\mathbf{j}_c$ signal magnitude is normalized by the maximum in the total thermoelectric image for the Néel skyrmion. On top of the total images, their profiles along the white dased lines are shown.

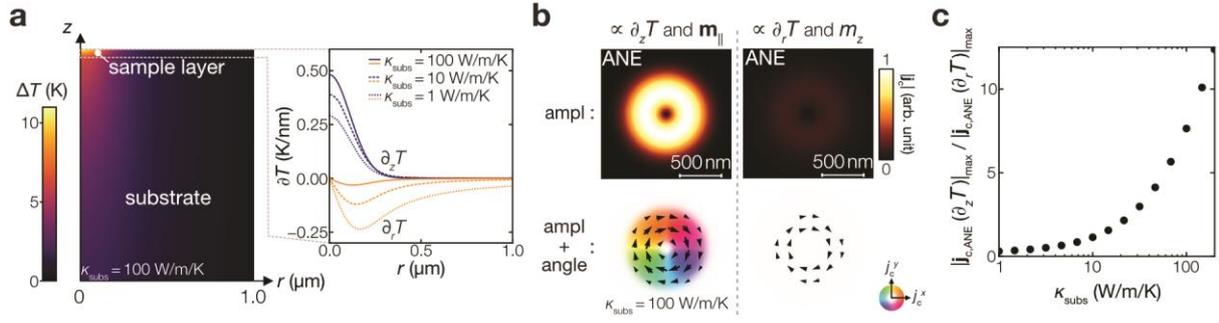

**Figure 5.** Temperature profiles and expected thermoelectric images at improved thermal conditions. (**a**) Calculated temperature profile for the substrate with $\kappa_{subs}$ = 100 Wm$^{-1}$K$^{-1}$. The left panel shows the temperature change $\Delta T$ around the sample layer. The right panel shows the out-of-plane (along the $z$ axis) and in-plane (along the $r$ axis) temperature gradients in the sample layer, where the values are averaged along the $z$ axis in the sample layer. (**b**) Resultant charge current signals at $\kappa_{subs}$ = 100 W m$^{-1}$K$^{-1}$. $\partial_z T$ induces $\mathbf{j}_c$ proportional to $|\mathbf{m}_\parallel|$, while $\partial_r T$ induces $\mathbf{j}_c$ reflecting $m_z$ as shown in Fig. 4c,d. The magnitude is normalized by the maximum amplitude. (**c**) Ratio of the maximum $|\mathbf{j}_{c,ANE}(\partial_z T)|$ to the maximum $|\mathbf{j}_{c,ANE}(\partial_r T)|$ as a function of $\kappa_{subs}$.

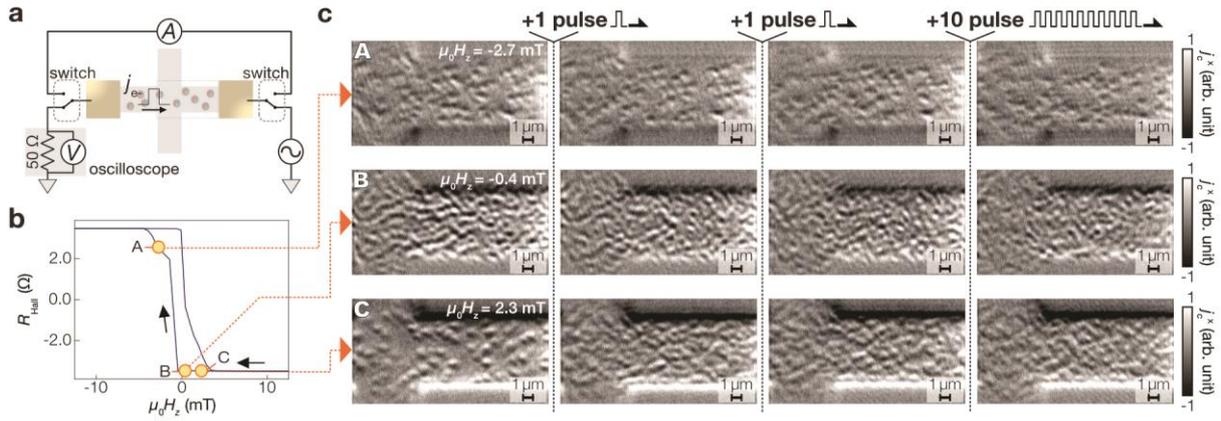

**Figure 6.** Thermoelectric microscopy with driving current pulses. (**a**) Schematic illustration of the experimental setup. $j_{e^-}$ denotes the flow of electrons in the Hall cross. The amplitude of the charge current was ~9 mA and the pulse duration was 100 ns, where the applied pulses are characterized using an oscilloscope inserted between the ground. (**b**) $H_z$ dependence of $R_{Hall}$. The thermoelectric images were measured at the $H_z$ values marked by yellow circles with alphabetical labels A-C. (**c**) Thermoelectric images after applying pulses. The images were recorded at the initial state just after $\mu_0 H_z$ is decreased from ~10 mT and at the states after applying 1, 1 (2 in total), and 10 (12 in total) pulses.

# Supplementary Information for
# Thermoelectric microscopy of magnetic skyrmions

Ryo Iguchi[1], Shinya Kasai[1,2], Kazushige Koshikawa[3], Norimichi Chinone[3], Shinsuke Suzuki[3], and Ken-ichi Uchida[1,4,5]

[1]National Institute for Materials Science, Tsukuba 305-0047, Japan.

[2]PRESTO, Japan Science and Technology Agency, Saitama 332-0012, Japan.

[3]Hamamatsu Photonics K.K., Hamamatsu 431-3196, Japan.

[4]Department of Mechanical Engineering, The University of Tokyo, Tokyo 113-8656, Japan.

[5]Center for Spintronics Research Network, Tohoku University, Sendai 980-8577, Japan.

Correspondence and requests for materials should be addressed to R.I. (e-mail: IGUCHI.Ryo@nims.go.jp)

**Optical performance of focused laser beam.**

Figure S1 shows the optical image of 0.5-µm-wide lines placed with 1 µm interval (0.5 µm gap). The image resolves the line and gap, indicating the resolution around 500 nm.

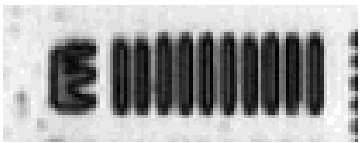

**Figure S1.** Optical image of 0.5-µm-wide lines with 1 µm interval.

**Magneto-optical imaging of magnetic texture.**

The appearance of the magnetic texture is confirmed by polar magneto-optical Kerr effect (MOKE) imaging of our sample as shown in Figure S2. A polarized white light beam was applied to the sample perpendicularly to the film plane and the reflected light distribution through an analyser is measured by an image sensor. In this configuration, the out-of-plane magnetic moment $m_z$ is detectable. The magnetic field is applied in the out-of-plane direction. Figure S3c shows the evolution of the MOKE images with the magnetic field magnitude $H_z$. Similarly to the thermoelectric images shown in Fig. 3, the magnetic texture appears at $\mu_0 H_z$ ~ -1.6 mT and the particle-like objects start to appear around $\mu_0 H_z$ ~ 3 mT. At $\mu_0 H_z$ = 4.0 mT, there can be found the isolated skyrmions, consistent with the thermoelectric imaging. By further increasing $H_z$, the magnetic texture disappears as the magnetisation is fully aligned along the applied field (Fig. S2b).

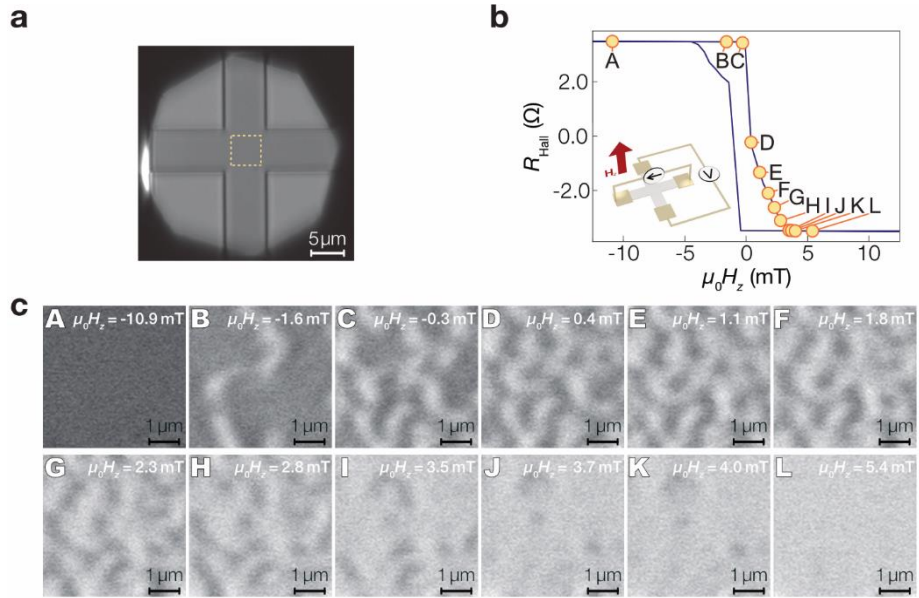

**Figure S2.** Magneto-optical imaging of magnetic texture. (**a**) Raw MOKE image measured at the magnetic field point A. (**b**) $H_z$ dependence of the Hall resistance $R_{Hall}$ of the sample system, which reflects the magnetic moment perpendicular to the plane. The MOKE images were measured at the magnetic field marked with yellow circles in the $R_{Hall}$-$H_z$ curve. (**c**) MOKE images at various values of $H_z$. The displayed area corresponds to the yellow rectangle area in Fig. S2a. The displayed images are the differential images from a background image taken at the field point A.

**Circumferential profile of thermoelectric image obtained by two-dimensional detection.**

Figure S3 shows the circumferential profiles of the magnetic circular objects in the amplitude image measured at the field label H in Fig. 3b. We calculated the profiles for three objects labelled α, β, and γ (Fig. S3b). The averaged profiles are calculated using the values in the annulus with the radius of the small (large) ring being about 230 nm (360 nm). As we discussed in the Results section of the main text (Simulation of thermoelectric images due to skyrmions), the Néel skyrmions are expected to show the uniform amplitude profile along the circumference while the trivial bubbles are expected to show the nonuniform or asymmetric profile (see Fig. 4f). Figure S3c shows, however, that the profiles do not have sufficient S/N ratio for experimental distinction.

**Figure S3.** Circumferential profile of magnetic objects in thermoelectric image. (**a**) Schematic and regions of the annulus used for the calculation. (**b**) Amplitude image of thermoelectric

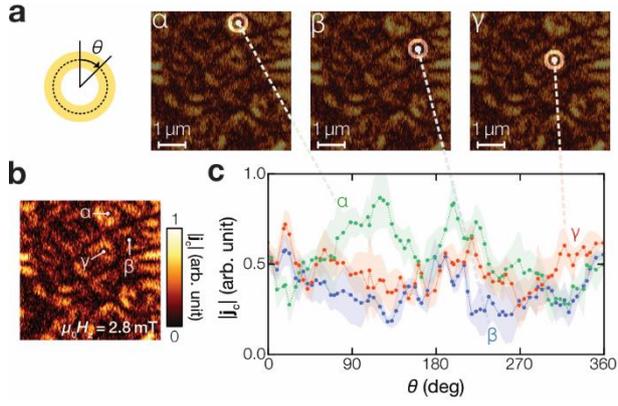

microscopy (the same as Fig. 3b, H). (**c**) Amplitude profile as a function of azimuthal angle $\theta$. The shadowed regions respectively cover the value range within the standard deviation.

**Skyrmion diameter dependence of the thermoelectric images.**

The diameter of the skyrmions is estimated by comparing the experimental and calculated results. Figure S4a shows the skyrmion diameter ($d$) dependence of the simulated thermoelectric images. By comparing the diameter of the ring-like pattern in the amplitude image at H of Fig. 3b and the calculated amplitude images, the skyrmions' diameter is found to be around 1000 nm. Interestingly, Fig. S4a indicates that although the magnitude decreases as $d$ decreases (see Fig. S4c), the spatial variation of the $\mathbf{j}_c$ direction around the skyrmion's centre remains even when the optical laser spot size is comparable to $d$ (Fig. S4b). This is due to the combination of steeply-changing magnetic structure of the skyrmions and spin-caloritronic phenomena. The sign of the thermoelectric signals is reversed between one and the other halves of each skymion. Thus, even if the heating diameter is comparable to the skyrmions, the shift of the heating centre from the skyrmion's centre ensures finite thermoelectric signals. We note that apart from the skyrmions' core, the spatial distribution is simply broadened following to the optical limitation.

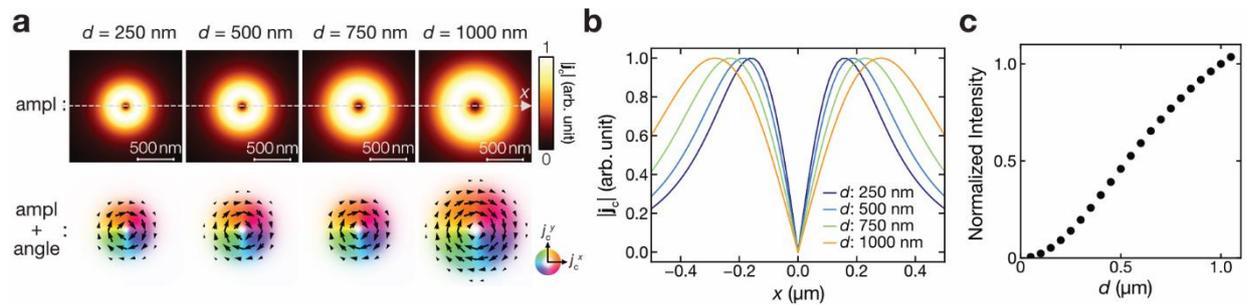

**Figure S4.** Temperature profiles and expected thermoelectric images. (**a**) skyrmion diameter $d$ dependence of the thermoelectric images. The signals are normalized in each image with the maximum value, which is plotted in Fig. S4c. (**b**) The profile of the total $|\mathbf{j}_c|$ image along the centre line for various values of $d$. (**c**) The maximum value of the total $|\mathbf{j}_c|$ images as a function of $d$, where the values are normalized at $d = 1.0$ μm.

**Optical image of prepared sample.**

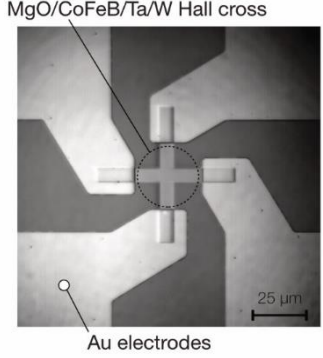

**Figure S5.** Optical image of the Hall cross and four Au electrodes used for two-dimensional thermoelectric current detection.

**Extraction of thermoelectric image due to spin caloritronic phenomena.**

The thermoelectric images shown in the main text were obtained after subtracting periodic environmental noise and background offset from the raw thermoelectric images. Firstly, the periodic noise with high spatial frequency (the wavy pattern in Fig. S6) was eliminated using a high-pass filter in the following procedure. The raw image was converted into a Fourier transformation (FT) image, the points with high wave vectors and large intensity were subtracted. Then, the FT image was converted back into the thermoelectric image. Secondly, the uniform background offset was subtracted, which is due to the electronic circuit, such as the characteristic offset of the amplifier and the thermoelectric current due to the environmental temperature difference inside the circuit. The offset value was calculated as the averaged value over the regions A-D shown in the leftmost images of Fig. S6, where A-D locate outside the Hall cross and no spin-caloritronic signals are expected. In the thermoelectric images after the offset subtraction, negative/positive signals can be found on the top/bottom of the horizontal section of the Hall cross. These signals might be due to the magnetic moments lying in-plane at the Hall cross edges and/or the asymmetric in-plane temperature gradient induced by edge illumination[1]; while at inside, the in-plane temperature gradient is radial and thus the signal is induced only when the out-of-plane magnetic moment $m_z$ distributes asymmetrically in the heating spot, at the edges, the in-plane temperature gradient is formed only toward the other edge and thus it simply reflects $m_\perp$. In addition, similar situation can happen for edges where the electrical sensitivity changes, *i.e.* around the boundary of the centre part of the Hall cross. We note that these parasitic signals do not affect the signals inside the cross area of the Hall cross, which we discussed in Fig. 3b.

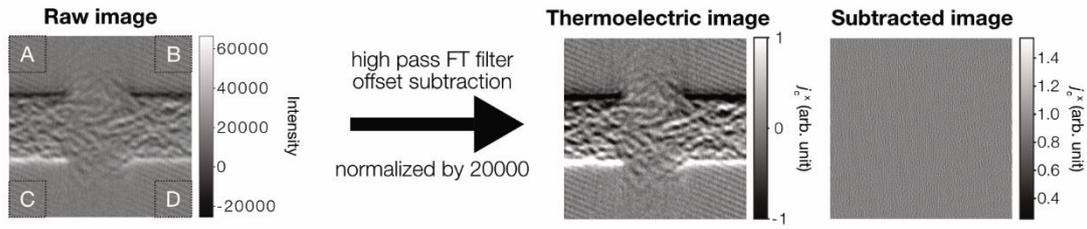

**Figure S6.** Schematic of signal processing. Raw, thermoelectric, and subtracted images are shown, where the subtracted image consists of the high frequency environmental noise and uniform offset. The raw image was recorded with 16-bit resolution for 32 times. The rectangles with the labels A, B, C, and D indicate the regions used for calculating the uniform offset.

**Power dependence of thermoelectric image.**

To check that the magnetic texture in our sample is not changed by the laser heating, we measured the laser power $P_{in}$ dependence of the thermoelectric images. Figure S7 shows that the distribution of the thermoelectric signals do not change when $P_{in} < 1.14$ mW, confirming that the experimental results shown in the main text, measured at $P_{in} = 0.84$ mW, are not affected by laser light effects[2]. Note that small change in the patchy pattern can be found at 1.28 and 1.43 mW irradiation.

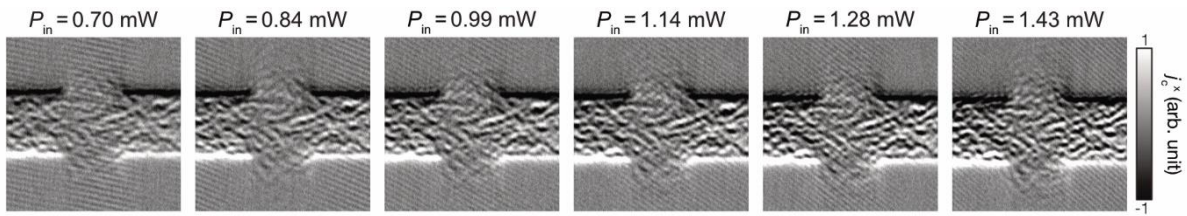

**Figure S7.** Irradiation power $P_{in}$ dependence of thermoelectric images.

**Estimation of the magnitude of the anomalous Nernst and spin Seebeck effects.**

In order to estimate the magnitude of the ANE and SSE contributions, we compared the magnitude of the transverse thermopower in the MgO/CoFeB/Ta/W multilayer film with the in-plane (IP) and out-of-plane (OP) magnetization[3,4]. For the IP (OP) magnetized system, both the ANE and SSE appear (only the ANE appears). The sample for this measurement was made on the thermally oxidized Si substrate, where the width $w$ is 2.0 mm and the length is 6.0 mm. The MgO/CoFeB/Ta/W film was deposited on the whole surface of the substrate and annealed in the same manner as the device used for the thermoelectric imaging. The sample was loaded on a heat bath and cramped by a heater block, which is thermally isolated from the environment so that the entire applied heater power $P_{\text{heater}}$ flows to the heat bath through the sample[5]. The length of the heater block $l'$ is 5.0 mm and the thermoelectric voltage $V$ induced within $l'$ was measured using a voltmeter. With applying $P_{\text{heater}} = 150$ mW to the heater block and the magnetic field to the sample, we measured the thermoelectric voltage along the length direction. The recorded voltage $V$ is converted into the transverse thermopower $S$ using the relation

$$S = \frac{V}{l'\nabla T} \tag{S1}$$

by estimating the temperature gradient $\nabla T$ applied to the sample. For the IP magnetized system, $\nabla T$ is simply determined by the thermal conductivity $\kappa$ of the sample layer:

$$\nabla T = \frac{P_{\text{heater}}}{\kappa l' w} \tag{S2}$$

For the OP magnetized system, $\nabla T$ is approximately determined by $\kappa_{\text{subs}}$ and thickness $t_{\text{subs}}$ of the substrate because the thermal resistance is governed by the substrate:

$$\nabla T = \frac{P_{\text{heater}}}{\kappa_{\text{subs}} l' t_{\text{subs}}} \tag{S3}$$

Figure S8 shows the magnetic field magnitude $H$ dependence of $S$ for the IP and OP magnetized samples. As the ANE and SSE change the sign of the output voltages when the magnetization is reversed, we extracted $S_{\text{IP}}$ and $S_{\text{OP}}$ based on the differences of the averaged $S$ at the positive and negative fields, where $\mu_0|H|$ is larger than 200 mT with $\mu_0$ being the vacuum permeability. Then, the ANE- and SSE-induced thermopowers ($S_{\text{ANE}}$ and $S_{\text{SSE}}$) are calculated by the following relations

$$S_{\text{ANE}} = S_{\text{OP}} \tag{S4}$$

$$S_{\text{SSE}} = S_{\text{IP}} - S_{\text{OP}} \tag{S5}$$

The obtained values are $S_{\text{ANE}} = 0.22$ μV/K and $S_{\text{SSE}} = -0.14$ μV/K, of which the sign and magnitude are consistent with previous reports[6,7]. For the above calculation, we assumed the thermal conductivities used for the temperature profile calculation in Fig. 4a.

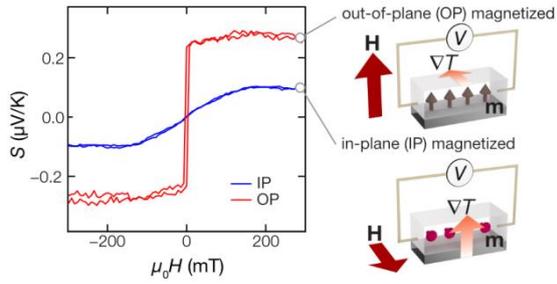

**Figure S8.** Magnetic field dependence of thermopower under uniform heating. The external field with the magnitude *H* was applied along the in-plane (IP) and out-of-plane (OP) directions of the multilayer on the rectangular-shaped substrate.